\documentclass{article}

\usepackage{graphicx}

\usepackage{amsmath}

\begin{document}

\title{Dirac bracket and time dependent constraints}

\author{N Barros e S\'a $^{1,2}$}

\date{\small{$^1$ FCT, Universidade dos A\c{c}ores, 9500-321 Ponta Delgada, Portugal\\
$^2$ Instituto de Astrof\'{\i}sica e Ci\^{e}ncias do Espa\c{c}o, 1349-018
Lisboa, Portugal}}

\maketitle

{\abstract{We provide a simultaneous derivation of the Dirac bracket and of the equations of motion for second-class constrained systems when the constraints are time-dependent. The necessity of time-dependent gauge-fixing conditions is shown in the example of parameterized mechanics and is illustrated geometrically.}}

\section{Introduction}

Constrained systems play a major role in modern physics, most notably in
gauge theories and in general relativity. Dirac established a theory %MDPI EE: Please check intended meaning has been retained
 for the canonical quantization of constrained
systems \cite{dir2}, introducing a new bracket, known today as a Dirac bracket, which
satisfies all of the desirable properties for a Poisson bracket (linearity in both
arguments, Leibnitz rule, antisymmetry, and the Jacobi identity), and which
preserves the form of Hamilton's equations of motion when written in terms
of brackets.

In the case of pure gauge theories, gauge-fixing conditions which are time-independent can be set, but in the case of diffeomorphism-invariant
theories, such as general relativity, the gauge-fixing conditions have to be time-dependent.

In Dirac's original work, the Hamiltonian %MDPI EE: Please check intended meaning has been retained
 and the constraints
are time-independent. The straightforward extension of Dirac's method to the case of time-dependent constraints was developed in Ref. \cite{muk}. Meanwhile, other approaches to this problem have been developed, such as the extension of phase space to include time as a canonical variable~\cite{cit1,cit2,cit4}, the modern differential geometric approaches of Refs. \cite{cit5,cit6}, and further applications~\cite{cit7}.

Here, we provide an alternative derivation of the expression for Dirac brackets, and one which concomitantly provides an
alternative route to solve %MDPI EE: Please check intended meaning has been retained
 the problem of time-dependent constraints provided in Ref. \cite{muk}. We also geometrically illustrate the need for time-dependent gauge-fixing conditions in theories which are invariant under time reparameterizations. 

%%%%%%%%%%%%%%%%%%%%%%%%%%%%%%%%%%%%%%%%%%
\section{Gauge-Fixed Constrained Systems}\label{s1}

In this section, we will not review Dirac's method, which may already be familiar %MDPI EE: Please check intended meaning has been retained
 to
many readers, and which is very clearly outlined %MDPI EE: Please check intended meaning has been retained
 in Dirac's %MDPI EE: Please check intended meaning has been retained
  original work \cite%
{dir2} and in many other works~\cite{cit1,hen,rot,sund,der1,itz}; rather, we will start from a
Hamiltonian which is already in the form%
\begin{equation}
H=H_{0}+\sum_{a=1}^{2M}u_{a}c_{a},\label{prie}
\end{equation}%
arising from Dirac's procedure, and where to each first-class constraint a
``gauge-fixing'' constraint has been added by hand, with the whole set of
constraints becoming second-class.

The antisymmetric matrix%
\begin{equation}
M_{ab}=\left\{ c_{a},c_{b}\right\}
\end{equation}%
is, by definition of second-class property, non-singular,%
\begin{equation}
\det M\neq 0.
\end{equation}

Ideally, for singular systems, one would like to be able to perform a canonical transformation to a set of phase space variables containing as a subset canonical pairs which are unconstrained and provide a representation of reduced phase space \cite{sund}; that is, which form a set of physical variables that capture the physical sector of the theory \cite{der1,der2} with unambiguous dynamics. However, not only may such a transformation not exist globally, but also, %MDPI EE: Please check intended meaning has been retained
 in general, one will not be able to isolate the variables of reduced phase space. Therefore, one may resort%MDPI EE: Please check intended meaning has been retained
  to gauge-fixing; that is, to add as many extra constraints as there are first-class constraints, such that the whole set of constraints becomes second-class. This procedure may still be of limited validity due to Gribov obstructions \cite{hen}, but this is what we assume has been performed %MDPI EE: Please check intended meaning has been retained
   in Equation \eqref{prie}, as we are specifically interested in the study of time-dependent constraints.

Dirac showed that requiring that time evolution preserves the second-class constraints one can solve for the $u_{a}$ in Equation \eqref{prie} and set the second-class constraints strongly to zero, and that the time evolution of a phase space variable is then given by%
\begin{equation}
\dot{F}=\left\{ F,H_{0}\right\} _{D}+\frac{\partial F}{\partial t},  \label{hama}
\end{equation}
where (summation over repeated indices is understood)
\begin{equation}
\left\{ F,G\right\} _{D}=\left\{ F,G\right\} -\left\{ F,c_{a}\right\}
M_{ab}^{-1}\left\{ c_{b},G\right\} \label{dbra}
\end{equation}%
is the Dirac bracket.

The straightforward extension of Dirac's method to time-dependent constraints was performed %MDPI EE: Please check intended meaning has been retained
 in Ref. \cite{muk}, where it was shown that Equation \eqref{hama} should be replaced by%
\begin{equation}
\dot{F}=\left\{ F,H_{0}\right\} _{D}+\left( \frac{\partial F}{\partial t}%
\right) _{D},  \label{eno0}
\end{equation}%
with
\begin{equation}
\left( \frac{\partial F}{\partial t}\right) _{D}=\frac{\partial F}{\partial t%
}-\left\{ F,c_{a}\right\} M_{ab}^{-1}\frac{\partial c_{b}}{\partial t}.  \label{eno}
\end{equation}

%%%%%%%%%%%%%%%%%%%%%%%%%%%%%%%%%%%%%%%%%%
\section{Parameterized Mechanics}

We consider the example of a theory describable by an action principle, from
an action%
\begin{equation}
S=\int_{t_{1}}^{t_{2}}\mathcal{L}\left( x^{\mu },\dot{x}^{\mu }\right) dt
\label{act}
\end{equation}%
which is invariant under redefinitions of the evolution parameter,%
\begin{equation}
t\rightarrow f\left( t\right) .  \label{par}
\end{equation}%
This %MDPI: please confirm if the noindent format should be kept in the full text
 is a diffeomorphism-invariant theory in one dimension. We use the
calligraphic letter $\mathcal{L}$ to denote the Lagrangian while utilizing %MDPI EE: Please check intended meaning has been retained
 the capital letter $L$ to denote another function, for reasons which will become
clear below.

The Lagrangian $\mathcal{L}$ cannot depend explicitly on $t$ and it must be
homogeneous of the first kind of the derivatives of the velocities \cite%
{dir2,sund},%
\begin{equation}
\mathcal{L}\left( x^{\mu },\lambda \dot{x}^{\mu }\right) =\lambda \mathcal{L}%
\left( x^{\mu },\dot{x}^{\mu }\right),
\end{equation}%
with $\mu =0,\ldots ,d$. Moreover, if we assume that one of the
configuration space variables, $x^{0}$, is a monotonous function of time, $%
dx^{0}/dt>0$, this condition can be stated as%
\begin{equation}
\mathcal{L}\left( x^{\mu },\dot{x}^{\mu }\right) =L\left( x^{0},x^{i},\frac{%
\dot{x}^{i}}{\dot{x}^{0}}\right) \dot{x}^{0},
\end{equation}%
with $i=1,\ldots ,d$. The momenta conjugate to the configuration space
variables are%
\begin{eqnarray}
p^{0} &=&L-\frac{\partial L}{\partial \left( \dot{x}^{i}/\dot{x}^{0}\right) }%
\frac{\dot{x}^{i}}{\dot{x}^{0}}  \label{pi1} \\
p^{i} &=&\frac{\partial L}{\partial \left( \dot{x}^{i}/\dot{x}^{0}\right) }%
,  \label{pi2}
\end{eqnarray}%
and the Hamiltonian vanishes, %MDPI EE: Please check intended meaning has been retained

\begin{equation}
\mathcal{H}=p^{0}\dot{x}^{0}+p^{i}\dot{x}^{i}-\mathcal{L}=0.
\end{equation}

Assuming that the relation \eqref{pi2} between the $p^{i}$ and the $\dot{x}%
^{i}/\dot{x}^{0}$ is invertible (if it is not, then further constraints show up---but we assume it is, in order to concentrate on the reparameterization invariance alone), one can write Equation \eqref{pi1} in the
form%
\begin{equation}
p^{0}+H\left( x^{0},x^{i},p^{i}\right) =0,  \label{coco}
\end{equation}%
with%
\begin{equation}
H=\frac{\partial L}{\partial \left( \dot{x}^{i}/\dot{x}^{0}\right) }\frac{%
\dot{x}^{i}}{\dot{x}^{0}}-L ,
\end{equation}%
which is clearly recognizable as the Legendre transform of $L$ in the
variables $\dot{x}^{i}/\dot{x}^{0}$; that is, the Hamiltonian associated
with the function $L$, seen as a Lagrangian. This is why we reserved the
calligraphic letters $\mathcal{L}$ and $\mathcal{H}$ for the original
Lagrangian and Hamiltonian of the theory.

Equation \eqref{coco} is a constraint, and because we assumed \eqref{pi2} to be
invertible, it is the only constraint, and hence a first-order one. Therefore,
the time evolution is undetermined, which we should have expected because we have
the freedom to choose any parameterization that we want for the evolution
variable $t$.

Now, we shall resort to gauge-fixing, as indicated in Section \ref{s1}. Adding the constraint $x^{0}=0$ seems to be a good
choice because it forms a second-class pair with the constraint \eqref{coco}: $\left\{ x^{0},p^{0}+H\left( x^{0},x^{i},p^{i}\right)
\right\} =1$. However, the Hamiltonian vanishes strongly, $\mathcal{H}=0$,
and using Equation \eqref{hama}, one would obtain %MDPI EE: Please check intended meaning has been retained
 $\dot{F}\left( x^{\mu },p^{\mu
}\right) =0$ for any function in phase space; there would be no dynamics! Of course, $x^{0}=0$ is incompatible with our assumption that $dx^{0}/dt>0$, but the same problem would arise for any constraint of the form $f(x^0,p^0,x^i,p^i)=0$, such that $\left\{ f,p^{0}+H\right\} \neq 0$.

The point is that the transformation (\ref{par}) that leaves the action 
\eqref{act} invariant is not a pure gauge transformation in the sense that the transformation is not only among the phase space variables; rather, it involves the evolution parameter, %MDPI EE: Please check intended meaning has been retained
 while for pure gauge
transformations, gauge-fixing conditions that involve only phase space
variables can be used. In the case of the invariance under
reparameterizations of the evolution variable, the ``gauge-fixing'' conditions
must necessarily involve the evolution variable itself, since that is the
variable that must be fixed.

\begin{figure}[tbp]
\centering
\includegraphics[width=.9\textwidth]{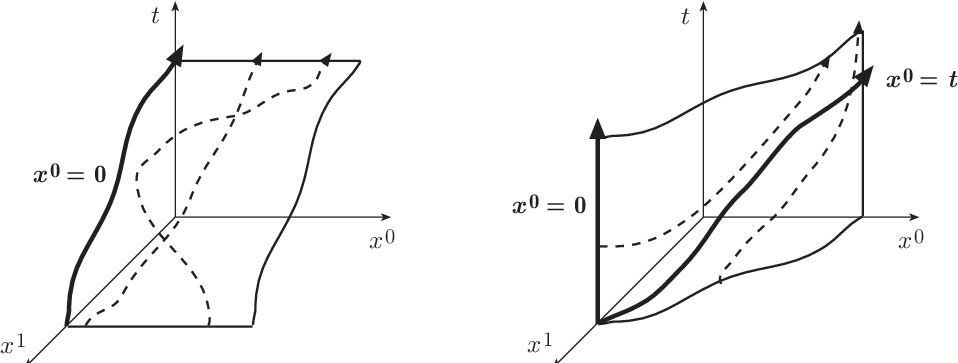}
\caption{\textbf{Left}: equivalent %MDPI: Please confirm whether an explanation of the arrow and dash line needs to be added to the figure caption.
 orbits when $x^{0}$ is pure gauge (dotted oriented lines). The
gauge can be fixed by setting $x^{0}=0$, thus selecting one of the orbits (the solid oriented line). \textbf{Right}: equivalent orbits for
invariance under reparameterization of the evolution variable $t$ (dotted oriented lines). Setting $x^{0}=t$ selects one of the orbits (solid oriented line on the right-hand side) and is acceptable,
but setting $x^{0}=0$ would freeze the dynamics (the vertical solid oriented line, which is not an orbit).}
\label{fi}
\end{figure}

This is illustrated in Figure \ref{fi}. On the left hand side of the
figure is an example where the variable $x^{0}$ is pure gauge ($x^1$ is some other variable, added for illustrative purposes---it may be looked upon as the rest of the phase space); that is, the
system is invariant under $x^{0}\rightarrow x^{0}+\alpha $, and on the right
hand side of the figure, there is an example where the system is invariant under
reparameterizations $t\rightarrow f\left( t\right) $. In both cases,
evolution is undetermined, with dotted lines representing equivalent
solutions. In the first case, the gauge can be fixed with $x^{0}=0$. Then, the
solution is unique (solid line). In the second case, setting $x^{0}=0$
(first solid line) freezes the dynamics ($x^{1}=const.$), but setting $%
x^{0}=t$ preserves the dynamics (second solid line).

The examples in Figure \ref{fi} also illustrate why Equation \eqref{hama} must %MDPI EE: Please check intended meaning has been retained
 be
improved when the constraints are time-dependent. After eliminating the
variable $x^{0}$, one has%
\begin{equation}
\left( \frac{\partial F}{\partial t}\right) _{x^{1}=cont}=\frac{\partial F}{%
\partial x^{0}}\frac{dx^{0}}{dt}+\frac{\partial F}{\partial t}.
\end{equation}%
Setting $x^{0}=0$, we obtain %MDPI EE: Please check intended meaning has been retained
\begin{equation}
\left( \frac{\partial F}{\partial t}\right) _{x^{1}=cont}=\frac{\partial F}{%
\partial t}.
\end{equation}%
However, setting $x^{0}=t$ we obtain %MDPI EE: Please check intended meaning has been retained
\begin{equation}
\left( \frac{\partial F}{\partial t}\right) _{x^1=cont}=\frac{\partial F}{%
\partial x^{0}}+\frac{\partial F}{\partial t}.
\end{equation}%
It is therefore the term $\partial F/\partial t$ that must %MDPI EE: Please check intended meaning has been retained
 be
corrected in Equation \eqref{hama}.

Returning %MDPI EE: Please check intended meaning has been retained
 to our theory, adding instead the gauge-fixing constraint%
\begin{equation}
C\left( x^{0},x^{i},p^{i}\right) -t=0\label{gfi}
\end{equation}%
to \eqref{pi1}, with%
\begin{equation}
\left\{ C,p^{0}+H\right\} \neq 0 ,
\end{equation}%
one obtains, using (\ref{eno0}) and (\ref{eno}) instead of (\ref{hama}),
\begin{equation}
\dot{F}=\frac{\partial F}{\partial t}+\frac{\left\{ F,p^{0}+H\right\} }{%
\left\{ C,p^{0}+H\right\} }.  \label{glo}
\end{equation}

As for the quantization of the theory, since $\mathcal{H}=0$, there is no
immediate Schr\"odinger picture, but the system can be described in the
Heisenberg picture because one can compute $\dot{F}$ using Equation \eqref{glo}.

Furthermore, after some manipulation, one arrives, starting from Equation \eqref{glo}, at
\begin{equation}
\dot{F}=\left\{ F,H\right\} _{D}+\frac{\partial F}{\partial t}+\left\{
F,p^{0}\right\} +\frac{\left\{ C,F\right\} \left\{ p^{0},H\right\} }{\left\{
C,p^{0}+H\right\} }.  \label{gla}
\end{equation}%
Hence, a Schr\"odinger picture can be recovered with the Hamiltonian given by $H$
 once one solves the constraints; %MDPI EE: Please check intended meaning has been retained
 the last three terms in this equation can
be interpreted as the explicit variation with respect to time. This is clear
if one chooses $C=x^{0}$ and solves the constraints for $x^{0}$ and $p^{0}$.
Then, Equation \eqref{gla} reads as follows:%
\begin{equation}
\dot{F}=\left\{ F,H\right\} _{D}+\frac{\partial F}{\partial t}+\frac{%
\partial F}{\partial x^{0}}-\frac{\partial F}{\partial p^{0}}\frac{\partial H%
}{\partial x^{0}} .
\end{equation}

The issue of recoverability of a Schr\"odinger picture in the general case of time-dependent constraints is discussed in Refs. \cite{cit3,cit8,cit9}.

The action \eqref{act} can be obtained by the inverse procedure, starting
from%
\begin{equation}
S=\int_{x_{0(1)}}^{x_{0(2)}}L\left( x^{0},x^{i},\frac{dx^{i}}{dx^{0}}\right)
dx^{0}
\end{equation}
and parameterizing the variable $x^{0}=x^{0}\left( t\right) $ \cite%
{dir2,sund}. That is why it is called parameterized~mechanics.

%%%%%%%%%%%%%%%%%%%%%%%%%%%%%%%%%%%%%%%%%%
\section{General Relativity}

General relativity is a field theory, and as such, its treatment is much more elaborate
than that of mechanics, which has a discrete number of degrees of freedom. We
will not review the extension of analytical mechanics to field theories, which is available elsewhere in many
textbooks \cite{hen,rot,sund,itz}, nor the several complications that may arise when imposing gauge-fixing conditions, such as the existence or not of global gauge-fixing conditions, the presence of surface terms, etc., (for a review of many of these matters, see~\cite{adm} or \cite{ish}). The sole purpose of this section is to provide an important example where %MDPI EE: Please check intended meaning has been retained
time-dependent constraints may show up, as the imposition of time-dependent gauge-fixing conditions is one of several attempts at dealing with the Hamiltonian formulation and general relativity. %MDPI EE: Please check intended meaning has been retained

The action
 for general relativity can be written in the form
\begin{equation}
S=\int dt \int d^3 x\left( \frac{1}{2}\pi^{ij} \dot{g}_{ij}-NH-N^iP_i\right) ,
\end{equation}
where $H$ and $P_i$ are functions of the space--space components of the metric $g_{ij}$
and their conjugate momenta $\pi^{ij}$, and $N$ and $N^i$ are functions that involve the
space--time and time--time components of the metric. Since the time derivatives
of these latest components of the metric do not show up in the action %MDPI EE: Please check intended meaning has been retained
, the
function $N$ and the vector $N^i$, called the lapse and the shift, respectively, act
as Lagrange multipliers for the constraints $H$ and $P_i$, called the
Hamiltonian constraint and the momentum constraint, respectively. This is the so-called
ADM decomposition of the action \cite{adm}.

Therefore, general relativity is a totally constrained theory, with its Hamiltonian
density being a combination of four constraints. These four constraints arise from the invariance under the four-dimensional (per spacetime point) group of diffeomorphisms in four dimensions, and they turn out to be first-class. Since the space--space part of the metric has six independent components, one arrives at the well-known counting of $6 -4 =2$ degrees of freedom per point for general relativity. One way to fix the gauge is to add four extra~constraints,

\begin{equation}
C^\alpha\left(x\right) = X^\alpha\left(\pi^{ij}(x),g_{ij}(x)\right)-x^\alpha = 0
\end{equation}
to the existing $H$ and $P_i$, such that, together, the eight of them form a set of second-class constraints. Here, $X^\alpha$ are some of the functions of the canonical variables at the point $x$. This can be interpreted as choosing a coordinate system \cite{dir2,adm,ish,and}, and it is the analogue of \mbox{Equation \eqref{gfi}}. Indeed, the original constraints of the theory arise precisely because of the freedom to choose the coordinate system, just like the constraint \eqref{coco} appears in parameterized mechanics because of the freedom to choose the evolution parameter.

\section{Analytical Mechanics in Matrix Form}

We shall use the notation in \cite{gold}, which we find more practical for
the present purpose. In this notation, the set of canonical coordinates $%
x_{i}$ and $p_{i}$ is composed into a column vector, and the same is performed %MDPI EE: Please check intended meaning has been retained
for the derivatives of a function $F$ with respect to the canonical~variables,
\begin{equation}
\eta =\left[
\begin{array}{c}
x_{1} \\
\cdots \\
x_{N} \\
p_{1} \\
\cdots \\
p_{N}%
\end{array}%
\right] \quad \quad \frac{\partial F}{\partial \eta }=\left[
\begin{array}{c}
\partial F/\partial x_{1} \\
\cdots \\
\partial F/\partial x_{N} \\
\partial F/\partial p_{1} \\
\cdots \\
\partial F/\partial p_{N}%
\end{array}%
\right].
\end{equation}

With the help of the matrix %MDPI: Please confirm if the bold formatting is necessary in all equations; if not, please remove it. 
\begin{equation}
\mathbf{J}=\left[
\begin{array}{cc}
0 & 1_{N\times N} \\
-1_{N\times N} & 0%
\end{array}%
\right],
\end{equation}%
Hamilton's equations of motion can be succinctly written as%
\begin{equation}
\dot{\eta}=\mathbf{J}\frac{\partial H}{\partial \eta },
\label{bla}
\end{equation}%
with $H$ being the Hamiltonian. The Poisson bracket between two functions $F$ and $%
G$ becomes%
\begin{equation}
\left\{ F,G\right\} =\left( \frac{\partial F}{\partial \eta }\right) ^{T}%
\mathbf{J}\frac{\partial G}{\partial \eta }
\end{equation}%
and the time evolution of an arbitrary function $F$ is%
\begin{equation}
\dot{F}=\left\{ F,H\right\} +\frac{\partial F}{\partial t}.
\end{equation}

\section{Time-Dependent Constraints}

Let us now admit that the Hamiltonian contains $2M$ second-class constraints
$c_{a}$, which we also write in the form of a column vector %
\begin{equation}
c\left( \eta ,t\right) =0.
\end{equation}%
Its time derivative is%
\begin{equation}
\dot{c}=\mathbf{C}\dot{\eta}+\frac{\partial c}{\partial t}=0.
\label{d1}
\end{equation}%
It must vanish, as the constraints are to be maintained with evolution. Here,
we have defined the $2M\times 2N$ matrix%
\begin{equation}
C_{ai}=\frac{\partial c_{a}}{\partial \eta _{i}}.
\end{equation}

Equation \eqref{d1} is a simple linear equation. It admits a solution if, and
only if \cite{adi}%
\begin{equation}
\left( \mathbf{1}-\mathbf{CC}^{-1}\right) \frac{\partial c}{\partial t}%
=0,  \label{cond}
\end{equation}%
where $\mathbf{C}^{-1}$ is a pseudoinverse of $\mathbf{C}$; that is, a $%
2N\times 2M$ matrix satisfying%
\begin{equation}
\mathbf{CC}^{-1}\mathbf{C=C}.
\end{equation}
Its solution is \cite{adi}%
\begin{equation}
\dot{\eta}=\left( \mathbf{1}-\mathbf{C}^{-1}\mathbf{C}\right) Z-\mathbf{C}%
^{-1}\frac{\partial c}{\partial t} ,\label{pri}
\end{equation}%
for some $Z$.

Demanding that Equation \eqref{pri} is applicable to all cases, even in the absence of constraints, when $c$ = 0 and $\mathbf{C}=0$, one must have
\begin{equation}
Z=\mathbf{J}\frac{\partial H}{\partial \eta },\label{ddd}
\end{equation}%
since we know that in the absence of constraints, Equation \eqref{bla} holds.
Hence,%
\begin{equation}
\dot{\eta}=\left( \mathbf{1}-\mathbf{C}^{-1}\mathbf{C}\right) \mathbf{J}%
\frac{\partial H}{\partial \eta }-\mathbf{C}^{-1}\frac{\partial c}{\partial t%
}.
\end{equation}

Now, we can compute the total derivative of a function $F\left( \eta
,t\right) $
\begin{eqnarray}
\dot{F} &=&\left( \frac{\partial F}{\partial \eta }\right) ^{T}\dot{\eta}+%
\frac{\partial F}{\partial t} = \nonumber \\
&=&\left( \frac{\partial F}{\partial \eta }\right) ^{T}\left( \mathbf{1}-%
\mathbf{C}^{-1}\mathbf{C}\right) \mathbf{J}\frac{\partial H}{\partial \eta }%
-\left( \frac{\partial F}{\partial \eta }\right) ^{T}\mathbf{C}^{-1}\frac{%
\partial c}{\partial t}+\frac{\partial F}{\partial t} = \nonumber \\
&=&\left( \frac{\partial F}{\partial \eta }\right) ^{T}\mathbf{J}\frac{%
\partial H}{\partial \eta }-\left( \frac{\partial F}{\partial \eta }\right)
^{T}\mathbf{C}^{-1}\mathbf{CJ}\frac{\partial H}{\partial \eta }+\frac{%
\partial F}{\partial t}-\left( \frac{\partial F}{\partial \eta }\right) ^{T}%
\mathbf{C}^{-1}\frac{\partial c}{\partial t}
\end{eqnarray}%
We want to rewrite this equation in the form%
\begin{equation}
\dot{F}=\left\{ F,H\right\} _{D}+\left( \frac{\partial F}{\partial t}\right)
_{D}
\end{equation}%
with a new ``Poisson bracket'' $\left\{ F,H\right\} _{D}$ that depends
linearly on $F$ and $H$, and a new ``partial derivative with respect to time''
$\left( \partial F/\partial t\right) _{D}$ which is linear in $F$. Then, we
must have%
\begin{eqnarray}
\left\{ F,G\right\} _{D} &=&\left\{ F,G\right\} -\left( \frac{\partial F}{%
\partial \eta }\right) ^{T}\mathbf{C}^{-1}\mathbf{CJ}\frac{\partial G}{%
\partial \eta }  \label{per1} \\
\left( \frac{\partial F}{\partial t}\right) _{D} &=&\frac{\partial F}{%
\partial t}-\left( \frac{\partial F}{\partial \eta }\right) ^{T}\mathbf{C}%
^{-1}\frac{\partial c}{\partial t}.  \label{per2}
\end{eqnarray}

We have not yet determined the pseudoinverse $\mathbf{C}^{-1}$ (recall that
the pseudoinverse is not unique), and neither have we showed that Equation \eqref{cond}
holds. Because $\mathbf{C}$ has a maximal rank (otherwise the constraints
would not be independent), $\mathbf{CC}^{T}$ is invertible and $\mathbf{C}%
^{T}\left( \mathbf{CC}^{T}\right) ^{-1}$ is a pseudoinverse (in fact, a
right pseudoinverse, and the Moore--Penrose pseudoinverse). This shows that
Equation \eqref{cond} holds. However, using this pseudoinverse would make the
Dirac bracket fail the requirement of antisymmetry in the exchange of $F$
with $G$. As we shall see, imposing this further requirement singles out one
pseudoinverse.

The first parcel on the right hand side of Equation \eqref{per1} is certainly antisymmetric because it is
the Poisson bracket. It is therefore enough to require antisymmetry for the
second parcel,
\begin{eqnarray}
\left( \frac{\partial F}{\partial \eta }\right) ^{T}\mathbf{C}^{-1}\mathbf{CJ%
}\frac{\partial G}{\partial \eta } = -\left( \frac{\partial G}{\partial
\eta }\right) ^{T}\mathbf{C}^{-1}\mathbf{CJ}\frac{\partial F}{\partial \eta }%
=  \nonumber \\
= -\left[ \left( \frac{\partial G}{\partial \eta }\right) ^{T}\mathbf{C}^{-1}%
\mathbf{CJ}\frac{\partial F}{\partial \eta }\right] ^{T} = -\left( \frac{%
\partial F}{\partial \eta }\right) ^{T}\mathbf{J}^{T}\left( \mathbf{C}^{-1}%
\mathbf{C}\right) ^{T}\frac{\partial G}{\partial \eta } .
\end{eqnarray}%
This can only be satisfied for all $F$ and $G$ if (notice that $\mathbf{J}%
^{T}=-\mathbf{J}$)
\begin{eqnarray}
\mathbf{C}^{-1}\mathbf{CJ} =\mathbf{J}\left( \mathbf{C}^{-1}\mathbf{C}%
\right) ^{T}\quad \Rightarrow   \nonumber \\
\Rightarrow \quad \mathbf{C}^{-1}\mathbf{CJC}^{T} =\mathbf{JC}^{T}\left( \mathbf{C}%
^{-1}\right) ^{T}\mathbf{C}^{T}=\mathbf{J}\left( \mathbf{CC}^{-1}\mathbf{C}%
\right) ^{T}=\mathbf{JC}^{T} .
\end{eqnarray}%
Since we know that%
\begin{equation}
\mathbf{M}=\mathbf{CJC}^{T}=\frac{\partial c_{a}}{\partial \eta _{i}}J^{ij}%
\frac{\partial c_{b}}{\partial \eta _{j}}=\sum_{i=1}^{N}\frac{\partial c_{a}%
}{\partial x_{i}}\frac{\partial c_{b}}{\partial p_{i}}-\frac{\partial c_{a}}{%
\partial p_{i}}\frac{\partial c_{b}}{\partial x_{i}}
\end{equation}%
is invertible \cite{dir2}, we finally obtain %MDPI EE: Please check intended meaning has been retained
\begin{equation}
\mathbf{C}^{-1}=\mathbf{JC}^{T}\mathbf{M}^{-1} .
\end{equation}%
Substitution into Equations \eqref{per1} and \eqref{per2} yields%
\begin{eqnarray}
\left\{ F,G\right\} _{D} &=&\left\{ F,G\right\} -\left( \frac{\partial F}{%
\partial \eta }\right) ^{T}\mathbf{JC}^{T}\mathbf{M}^{-1}\mathbf{CJ}\frac{%
\partial G}{\partial \eta } \\
\left( \frac{\partial F}{\partial t}\right) _{D} &=&\frac{\partial F}{%
\partial t}-\left( \frac{\partial F}{\partial \eta }\right) ^{T}\mathbf{JC}%
^{T}\mathbf{M}^{-1}\frac{\partial c}{\partial t} .
\end{eqnarray}

Or, back to indices notation,%
\begin{eqnarray}
\left\{ F,G\right\} _{D} &=&\left\{ F,G\right\} -\frac{\partial F}{\partial
\eta _{i}}J_{ij}\frac{\partial c_{a}}{\partial \eta _{j}}\left[ \frac{%
\partial c_{a}}{\partial \eta _{m}}J_{mn}\frac{\partial c_{b}}{\partial \eta
_{n}}\right] ^{-1}\frac{\partial c_{b}}{\partial \eta _{k}}J_{kl}\frac{%
\partial G}{\partial \eta _{l}}=  \nonumber \\
&=&\left\{ F,H\right\} -\left\{ F,c_{a}\right\} \left\{ c_{a},c_{b}\right\}
^{-1}\left\{ c_{b},G\right\}  \\
\left( \frac{\partial F}{\partial t}\right) _{D} &=&\frac{\partial F}{%
\partial t}-\frac{\partial F}{\partial \eta _{i}}J_{ij}\frac{\partial c_{a}}{%
\partial \eta _{j}}\left[ \frac{\partial c_{a}}{\partial \eta _{m}}J_{mn}%
\frac{\partial c_{b}}{\partial \eta _{n}}\right] ^{-1}\frac{\partial c_{b}}{%
\partial t}=  \nonumber \\
&=&\frac{\partial F}{\partial t}-\left\{ F,c_{a}\right\} \left\{
c_{a},c_{b}\right\} ^{-1}\frac{\partial c_{b}}{\partial t} .
\end{eqnarray}%
To summarize our results, we showed that one can work on the constraint
surface, defined by $c_{a}=0$, provided that one replaces the Poisson
bracket with the Dirac bracket%
\begin{equation}
\left\{ F,G\right\} _{D}=\left\{ F,G\right\} -\left\{ F,c_{a}\right\}
M_{ab}^{-1}\left\{ c_{b},G\right\} ,  \label{diro}
\end{equation}%
with $M_{ab}=\left\{ c_{a},c_{b}\right\} $, and the partial derivative with
respect to time with%
\begin{equation}
\left( \frac{\partial F}{\partial t}\right) _{D}=\frac{\partial F}{\partial t%
}-\left\{ F,c_{a}\right\} M_{ab}^{-1}\frac{\partial c_{b}}{\partial t}
.  \label{muko}
\end{equation}

Therefore, we provided an alternative and simultaneous derivation of Equations \eqref{dbra} and \eqref{eno}.

%%%%%%%%%%%%%%%%%%%%%%%%%%%%%%%%%%%%%%%%%%
\section{Discussion}

Time-dependent gauge-fixing conditions are important in a number of situations; namely, in general relativity. The straightforward method of dealing with them in conjunction with the Dirac bracket that we studied in this paper is one possible way of tackling the problem, for which we provided a new and simultaneous derivation of the relevant results. Furthermore, it is helpful to understand the reasons why, and in which situations, one has to work with time-dependent constraints when they arise from gauge-fixing, for which we provided a simple pedagogical and graphical example.

The reader should not get the wrong impression that this derivation overlooked %MDPI EE: Please check intended meaning has been retained
 dynamical aspects. In fact, it arises from the same principles used in standard textbooks, namely that the Hamiltonian $H$ generates the equations of motion, Equation \eqref{pri} together with \mbox{Equation \eqref{ddd}}, and the requirement of time preservation of the constraints, expressed in Equation \eqref{d1}.

\end{document}